\documentclass[a4paper,fleqn,usenatbib]{mnras} 

\usepackage{latexsym}
\usepackage{amssymb}
\usepackage{amsfonts}
\usepackage{amsmath}
\usepackage{bm}
\usepackage{graphicx}
\usepackage{subfigure}
\usepackage{times}
\usepackage{units}
\usepackage{hyperref}
\usepackage{multirow}
\usepackage{comment}
\usepackage{ulem}
\usepackage{hyperref}
\usepackage{graphicx}
\usepackage{acronym}
\usepackage{xcolor}
\usepackage{booktabs}

\usepackage{pifont} 

\usepackage{enumitem}
\setlist[itemize]{noitemsep, topsep=0pt}

\usepackage{multirow}
\usepackage{morefloats}
\usepackage{mathrsfs}
\usepackage{latexsym}
\usepackage{dcolumn}
\usepackage{lipsum}
\usepackage{mathtools}
\usepackage{cuted}

\renewcommand{\cite}[1]{\citep{#1}}

\newcommand{\GW}{GW170817}
\newcommand{\AT}{AT2017gfo} 
\newcommand{\GRB}{GRB170817A}

\newcommand{\mr}{mass ratio} 
\newcommand{\tlam}{reduced tidal parameter} 
\newcommand{\bnc}{bounce} 
\newcommand{\hajela}{\citep{2021arXiv210402070H}}

\newcommand{\surname}[1]{\text{#1}}

\newcommand{\ie}{\textit{i.e.}}
\newcommand{\eg}{\textit{e.g.}}

\newcommand{\be}{\begin{equation}}
\newcommand{\ee}{\end{equation}}
\newcommand{\bea}{\begin{eqnarray}}
\newcommand{\eea}{\end{eqnarray}}
\newcommand{\bel}{\begin{align}}
\newcommand{\eel}{\end{align}}

\def\md{M_{\rm ej}}

\def\vd{v_\infty}
\def\yd{Y_{e}}

\def\amd{\md}

\def\gccm{{\rm g\,cm^{-3}}}

\def\GMc2{{\rm G M_{\odot} c^{-2}}}

\def\nism{n_{\text{ISM}}}
\def\ccm{\,\text{cm}^{-3}}

\newacro{BH}{black hole}
\newacro{BBH}{binary black-hole}
\newacro{BHNS}{black-hole neutron-star}
\newacro{BNS}{binary neutron star}
\newacro{EM}{electromagnetic}
\newacro{EOS}{equation of state}
\newacroplural{EOS}[EOSs]{equations of state}
\newacro{GR}{general relativity}
\newacro{GRLES}{general-relativistic large-eddy simulation}
\newacro{GRHD}{general-relativistic hydrodynamics}
\newacro{GRMHD}{general-relativistic magnetohydrodynamics}
\newacro{GW}{gravitational wave}
\newacroplural{GW}[GWs]{gravitational waves}
\newacro{LES}{large-eddy simulation}
\newacroplural{LES}[LES]{large-eddy simulations}
\newacro{MRI}{magnetorotational instability}
\newacro{NR}{numerical relativity}
\newacro{NS}{neutron star}
\newacroplural{NS}[NSs]{neutron stars}
\newacro{SGRB}{short $\gamma$-ray burst}
\newacro{ISM}{interstellar medium}
\newacro{SED}{spectral energy distribution}
\newacro{EATS}{equal time arrival surface}
\newacroplural{EATS}[EATSs]{equal time arrival surfaces}
\newacro{RMS}{root-mean-square}
\newacro{LC}{light curve}
\newacroplural{LC}[LCs]{light curves}
\newacro{KH}{Kevin-Helmholtz}

\newacro{GRB}{$\gamma$-ray burst}
\newacro{ST}{Sedov-Taylor}
\newacro{BM}{Blandford-McKee}
\newacro{ODE}{ordinary differential equation}
\newacroplural{ODE}[ODEs]{ordinary differential equations}
\newacro{RK}{Runge–Kutta}

\renewcommand{\cite}[1]{\citep{#1}}
\title[Dynamical ejecta afterglow for \GRB{}]{
    Dynamical ejecta synchrotron emission as a 
    possible contributor to the changing behaviour of \GRB{} afterglow
}

\author[V. Nedora et al.]{
 Vsevolod \surname{Nedora}$^{1}$,
 David \surname{Radice}$^{2,3,4}$,
 Sebastiano \surname{Bernuzzi}$^{1}$,
 Albino \surname{Perego}$^{5,6}$,
 \newauthor\hspace{0.5mm}
 Boris \surname{Daszuta}$^{1}$,
 Andrea \surname{Endrizzi}$^{1}$, 
 Aviral \surname{Prakash}$^{2,3}$, and
 Federico \surname{Schianchi}$^{7}$
\\
  ${}^1$Theoretisch-Physikalisches Institut, Friedrich-SchillerUniversit\"{a}t Jena, 07743, Jena, Germany\\
  ${}^2$Institute for Gravitation \& the Cosmos, The Pennsylvania State University, University Park, PA 16802, USA\\
  ${}^3$Department of Physics, The Pennsylvania State University, University Park, PA 16802, USA\\
  ${}^4$Department of Astronomy \& Astrophysics, The Pennsylvania State University, University Park, PA 16802, USA\\
  ${}^5$Dipartimento di Fisica, Universit\`{a} di Trento, Via Sommarive 14, 38123 Trento, Italy\\
  ${}^6$INFN-TIFPA, Trento Institute for Fundamental Physics and Applications, via Sommarive 14, I-38123 Trento, Italy\\
  ${}^7$Institut f\"{u}r Physik und Astronomie, Universit\"{a}t Potsdam, Haus 28, Karl-Liebknecht-Str.  24/25,14476, Potsdam, Germany
}

\date{
  Accepted XXX. Received YYY; in original form ZZZ
}

\begin{document}
\label{firstpage}
\maketitle

\date{\today}

\begin{abstract}
    Over the past three years, the fading non-thermal emission from the \GW{} 
    remained generally consistent with the 
    afterglow powered by synchrotron radiation produced by the interaction of the 
    structured jet with the ambient medium.
    Recent observations by Hajela \textit{et al.}~2021 indicate the
    change in temporal and spectral behaviour in the X-ray band. 
    We show that the new observations are compatible with the 
    emergence of a new component due to 
    non-thermal emission from the fast tail of the
    dynamical ejecta of ab-initio \ac{BNS} merger simulations.
    This provides a new avenue to constrain binary parameters. 
    Specifically, we find that
    equal mass models with soft \ac{EOS} and 
    high \mr{} models with stiff \ac{EOS} are disfavored 
    as they typically predict afterglows that peak too early 
    to explain the recent observations.
    Moderate stiffness and mass ratio models, instead,
    tend to be in good overall agreement with the data.
\end{abstract}


\section{Introduction}
\label{sec:intro}

The \GW{} event marked the dawn of the era of multimessenger astronomy with compact binary mergers. 
This event was observed as \ac{GW} source, 
\GW~\citep{TheLIGOScientific:2017qsa,Abbott:2018wiz,LIGOScientific:2018mvr}; 
quasi-thermal \ac{EM} transient, commonly referred to as kilonova, \AT~\citep{Arcavi:2017xiz,Coulter:2017wya,Drout:2017ijr,Evans:2017mmy,Hallinan:2017woc,Kasliwal:2017ngb,Nicholl:2017ahq,Smartt:2017fuw,Soares-santos:2017lru,Tanvir:2017pws,Troja:2017nqp,Mooley:2018dlz,Ruan:2017bha,Lyman:2018qjg}; and \ac{SGRB}, \GRB{} \citep{Savchenko:2017ffs,Alexander:2017aly,Troja:2017nqp,Monitor:2017mdv,Nynka:2018vup,Hajela:2019mjy}, detected by the space observatories Fermi \citep{TheFermi-LAT:2015kwa} and INTEGRAL \citep{Winkler:2011}.

This \ac{SGRB} was dimmer then any other events of its class. 
Different interpretations for its dimness and slow rising flux were proposed: off-axis jet, cocoon or structured jet. 
Now it is commonly accepted that \GRB{} was a structured jet observed off-axis 
\citep[\eg][]{Fong:2017ekk,Troja:2017nqp,Margutti:2018xqd,Lamb:2017ych,Lamb:2018ohw,Ryan:2019fhz,Alexander:2018dcl,Mooley:2018dlz,Ghirlanda:2018uyx}.
The \GRB{} late emission, the afterglow, provided further information on 
the energetics of the event and on the properties of the circumburst medium \citep[\eg][]{Hajela:2019mjy}.

The non-thermal afterglow of \GRB{} has been observed for over three years,
fading after its peak emission at ${\sim}160$~days after merger. 
At the time of writing, $3.2\,$years past the merger, the post-jet-break 
afterglow 
is still being observed, albeit only in X-ray by Chandra \citep{2021arXiv210402070H} 
and in radio by VLA \citep{2021arXiv210304821B}, 
as its flux in optical wavelengths has decreased below the detection limit
\citep{Troja:2020pzf}. 
Up until $900\,$days after merger, the non-thermal emission in X-rays and radio have 
followed the typical post-jet-break afterglow decay, $t^{-p}$ \citep[][]{Sari:1997qe}. 
After $900\,$days, a flattening in the X-rays, \ie, 
behaviour divergent from the $t^{-p}$ decay, was observed \citep{Troja:2020pzf}.  
More recent observations by Chandra showed the emergence of a new 
rising component in X-ray, not accompanied by the increase in radio flux, 
indicating a change of the spectral behaviour of the afterglow
\citep{2021arXiv210402070H,2021arXiv210304821B}.

There are several possible explanations for this behaviour, that generally 
fall into two categories \citep{2021arXiv210402070H}.
The first one is related changes occurring in the same shock that produced 
the previously observed afterglow emission 
\citep{Frail:1999hk,Piran:2004ba,Sironi:2013tva,Granot:2017gwa,Nakar:2019fza} 
and include 
the transition of the blast wave to the Newtonian regime, 
energy injection into the blast wave, 
change of the \ac{ISM} density,
emergence of a counter-jet emission, 
and evolution of the microphysical parameters of the shock.
The second category tight to the emergence of a new emitting component
\citep[\eg]{Nakar:2011cw,Piran:2012wd,Hotokezaka:2015eja,Radice:2018pdn,Hotokezaka:2018gmo,Kathirgamaraju:2018mac,Desai:2018rbc,Nathanail:2020hkx,2021arXiv210404433I} and includes 
afterglow from the decelerating ejecta, 
produced at merger or/and after, 
and emission powered by accretion onto a newly formed compact object.
Notably, while the fall-back accretion scenario provides a tentative explanation for 
the X-ray excess in the observed spectrum, it requires a suppression mechanism at earlier times,
\eg, supression of the fall-back due to $r$-process heating \citep[\eg][]{Desai:2018rbc,2021arXiv210404433I},
as the earlier emission from \GW{} was consistent with the structured off-axis jet afterglow
\citep[\eg]{Troja:2020pzf,Hajela:2019mjy,2021arXiv210402070H}.
The kilonova afterglow, on the other hand, is a more straightforward explanation, 
as the kilonova itself has been observed and the emergence of its afterglow is only natural. 
Additionally, the X-ray excess, or in other words, stepper electron spectrum with lower 
$p$, is expected for non-relativistic outflows \citep{Kathirgamaraju:2018mac,Bell:1978,Blandford:1978,Blandford:1987}. 
In any case, this opens a new avenue for the multimessenger study of \GW{}.

In the past few years \GW{} and its \ac{EM} counterparts have been the subjects of intense 
investigations and the privileged target for numerical and analytical studies. 
The wealth of \ac{GW} models and analysis techniques 
allowed to constrain
the intrinsic parameters of the binary, such as the 
masses of the merged objects, and the properties of the \ac{EOS} of cold, beta-equilibrated nuclear matter.
The modeling of the kilonova \acp{LC} and spectra shed new 
light on the origin of the heaviest elements in the Universe, including 
lanthanides and actinides \citep{Barnes:2016umi,Kasen:2017sxr,Tanaka:2017qxj,Miller:2019dpt,Bulla:2019muo}, 
and constrained the properties of the matter ejected during the merger, 
\citep[\eg][]{Villar:2017wcc,Perego:2017wtu,Siegel:2019mlp,2021arXiv210101201B,2021arXiv210202229N}. 
The joint analysis of \ac{GW} and kilonova emission allowed to 
further constrain the properties of the binary and of the \ac{NS} \ac{EOS} 
\citep{Margalit:2017dij,Bauswein:2017vtn,Radice:2017lry,Dietrich:2018upm,Radice:2018ozg,Dietrich:2020efo,2021arXiv210101201B}.

It has long been suggested that the dynamical ejecta from \ac{BNS} mergers
can be the source of a non-thermal, synchrotron emission 
resulting from the interaction of these ejecta with the surrounding \ac{ISM}
\citep[\eg][]{Nakar:2011cw,Piran:2012wd,Hotokezaka:2015eja}.
Depending on the binary and \ac{EOS} properties, \ac{NR} simulations show the presence of fast,
mildly relativistic material at the forefront of the dynamical ejecta which can efficiently power this emission.
Thus, a non-thermal kilonova afterglow is expected as one of the possible observable \ac{EM} counterparts of \GW{} \citep[\eg][]{Hotokezaka:2018gmo,Radice:2018pdn,Radice:2018ghv,Kathirgamaraju:2019xwu}.

\begin{figure*}
    \centering 
    \includegraphics[width=0.329\textwidth]{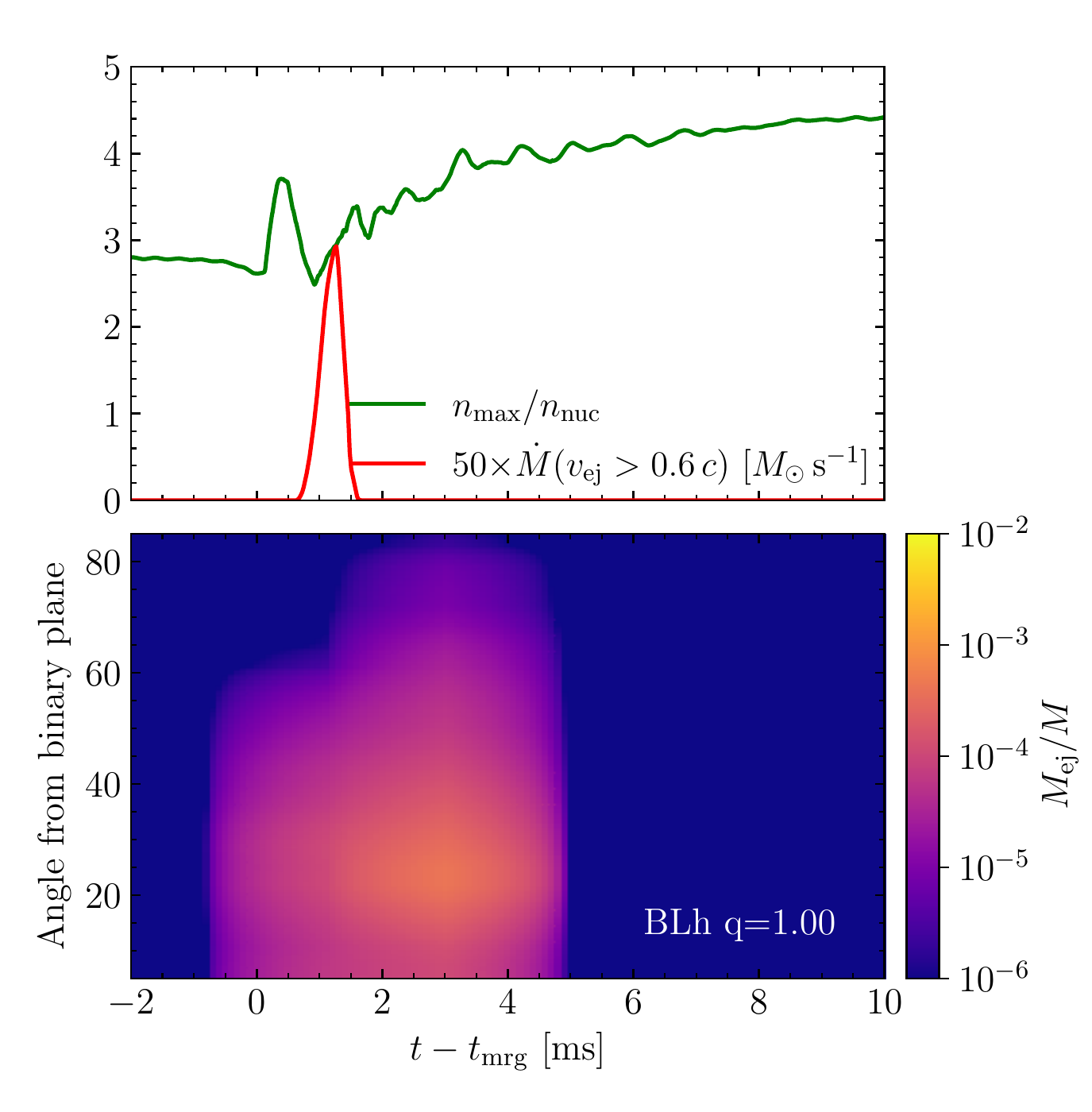}
    \includegraphics[width=0.329\textwidth]{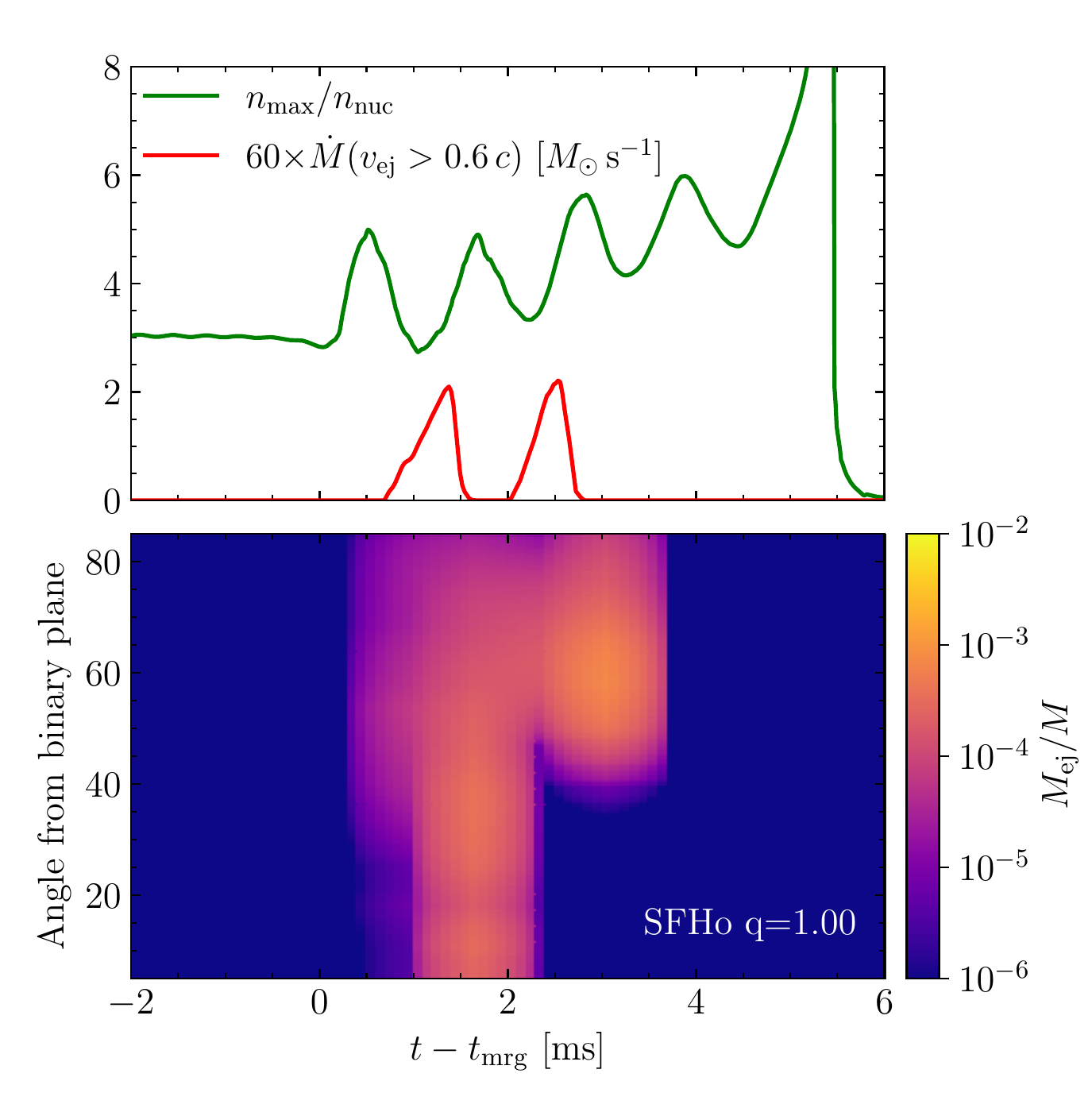}
    \includegraphics[width=0.329\textwidth]{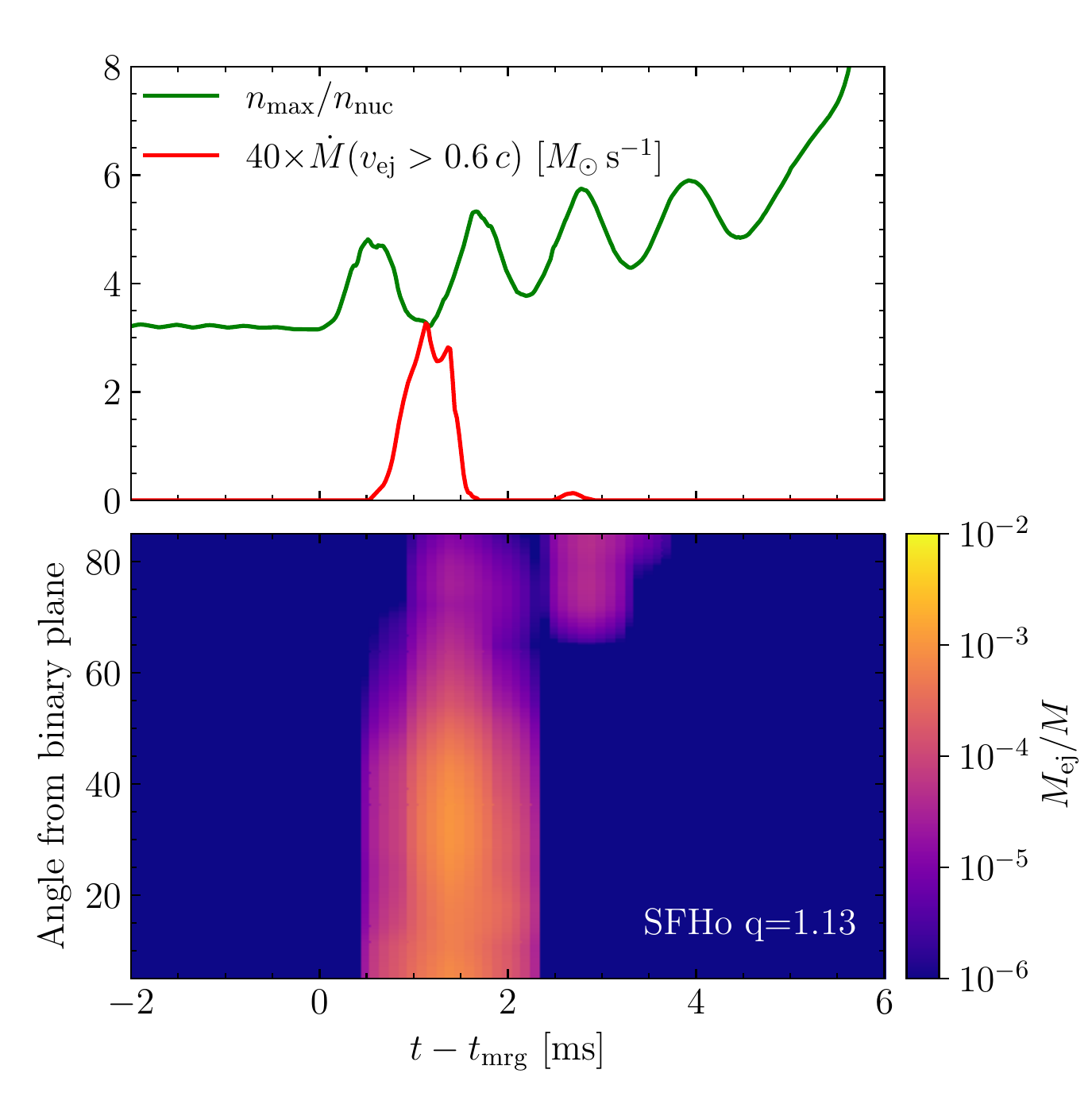}
    \caption{
        Ejection mechanism and properties of the fast tail of the ejecta shown for three simulations,
        with two \acp{EOS}: BLh and SFHo and two \mr{}s: $q=1.00$ and $q=1.22$.
        The upper panel in each plot shows the time evolution of the maximum density 
        in the simulation (green curves) and the mass flux of the ejecta with 
        asymptotic velocities exceeding $0.6c$ (red curve).
        The bottom panel shows the mass histogram of the fast ejecta tail as a function 
        of time. 
        In both panels the outflow rate and histograms are computed at a radius of 
        $R = 443$~km and shifted in time by 
        $R\langle\upsilon_{\rm fast}\rangle ^{-1}$, 
        $\langle\upsilon_{\rm fast}\rangle$ being the mass averaged velocity of 
        the fast tail at the radius $R$.
        The plot shows that most of the fast ejecta are generally produced at first core \bnc{} with a contribution from the second in models with soft \acp{EOS}.
    } 
    \label{fig:ejecta_v06_mech}
\end{figure*}

In this work we show that the observed X-ray behavior can be 
explained by an emerging non-thermal afterglow emission from the fast tail of \ac{BNS} dynamical ejecta.
We consider state-of-the-art \ac{NR} simulations targeted to the \GW{} event 
(\ie, with \GW{} binary chirp mass) and documented in \citet{Perego:2019adq,Nedora:2019jhl,Bernuzzi:2020tgt,Nedora:2020pak}.
We compute synthetic \acp{LC} from the simulated ejecta using semi-analytical methods and show that the peak time and flux are consistent with the recent
observations from some of the models.
This provides a new avenue to constrain the binary parameters, suggesting that the equal mass models with very soft \ac{EOS} peak too early to be consistent with observed changing behaviour of the \GRB{} X-ray afterglow.


\section{Binary neutron star merger dynamics and dynamical mass ejection}
\label{sec:ejecta}

\ac{NR} simulations of \ac{BNS} mergers provided a quantitative 
picture of the merger dynamics, mass ejection mechanisms and remnant evolution
\citep[\eg][]{Shibata:2019wef,Radice:2020ddv,Bernuzzi:2020tgt}.

We consider a large set of \ac{NR} \ac{BNS} merger simulations targeted to
\GW{} \citep{Perego:2019adq,Endrizzi:2019trv,Nedora:2019jhl,Bernuzzi:2020txg,Nedora:2020pak}.
These simulations were performed with the GR hydrodynamics code \texttt{WhiskyTHC} 
\citep{Radice:2012cu,Radice:2013xpa,Radice:2013hxh,Radice:2015nva},
and included neutrino emission and absorption using the M0 method described in 
\citet{Radice:2016dwd} and \citet{Radice:2018pdn}, and turbulent viscosity
of magnetic origin via an effective subgrid scheme, as described in \citet{Radice:2017zta} and \citet{Radice:2020ids}.
The impact of viscosity on the dynamical ejecta properties was investigated in 
\citet{Radice:2018ghv}, \citet{Radice:2018pdn}, and \citet{Bernuzzi:2020txg}, while the importance of neutrinos 
for determining the ejecta properties was discussed in \citet{Radice:2018pdn} and \cite{Nedora:2020pak}, 
where it was shown that neutrino reabsorption increases the ejecta mass and velocity -- 
two main quantities for the kilonova afterglow.
All simulations were performed using finite temperature, composition dependent nuclear \acp{EOS}. 
In particular, we employed the following set of \acp{EOS} to bracket the present uncertainties: 
DD2 \citep{Typel:2009sy,Hempel:2009mc},
BLh \citep{Logoteta:2020yxf}, 
LS220 \citep{Lattimer:1991nc},
SLy4 \citep{Douchin:2001sv,daSilvaSchneider:2017jpg}, and
SFHo \citep{Steiner:2012rk}. 
Among them, DD2 is the stiffest (thus providing larger radii, larger tidal deformabilities 
and larger \ac{NS} maximum masses), while SFHo and SLy4 are the softest.

\begin{table}
\begin{center}
\caption{
Properties of the fast tail of the dynamical ejecta 
(that has velocity $\upsilon>0.6$) for a list of \ac{NR} 
simulations from \citet{Nedora:2020pak} for which this ejecta 
is found. Columns, from left to right, are:
the \ac{EOS}, \tlam{}, \mr{}, 
mass of the fast tail, its mass-averaged electron fraction, and 
velocity, and the \ac{RMS} half-opening angle around the binary plane.
Asterisk next to \ac{EOS} indicate a model with subgrid turbulence.
}
\label{tab:sim}
\begin{tabular}{l c c c c c c}
\hline
EOS & $\tilde{\Lambda}$ & $q$ & $M_{\rm ej}$ & $\langle Y_e  \rangle$ & $\langle \upsilon_{\infty}  \rangle$ & $\langle \theta_{\rm RMS} \rangle$ \\
  &   &   & $[M_{\odot}]$ &   & $[c]$ &  [deg] \\ 
\hline\hline
BLh* & 541 & 1.00 & $1.52 \times 10^{-6}$ & 0.25 & 0.63 & 59.55 \\
BLh & 541 & 1.00 & $2.53 \times 10^{-5}$ & 0.32 & 0.68 & 30.05 \\
BLh & 539 & 1.34 & $1.37 \times 10^{-6}$ & 0.28 & 0.62 & 40.73 \\
BLh* & 539 & 1.34 & $8.02 \times 10^{-7}$ & 0.20 & 0.63 & 18.05 \\
BLh & 540 & 1.43 & $1.19 \times 10^{-8}$ & 0.32 & 0.60 & 60.74 \\
BLh* & 543 & 1.54 & $1.22 \times 10^{-6}$ & 0.31 & 0.62 & 61.23 \\
BLh & 538 & 1.66 & $1.25 \times 10^{-6}$ & 0.32 & 0.62 & 51.40 \\
BLh* & 532 & 1.82 & $6.40 \times 10^{-7}$ & 0.36 & 0.74 & 67.44 \\ \hline
DD2* & 853 & 1.00 & $6.65 \times 10^{-6}$ & 0.28 & 0.63 & 12.80 \\
DD2 & 853 & 1.00 & $9.65 \times 10^{-7}$ & 0.28 & 0.63 & 23.27 \\
DD2* & 847 & 1.20 & $4.19 \times 10^{-7}$ & 0.24 & 0.61 & 20.02 \\
DD2 & 846 & 1.22 & $1.34 \times 10^{-5}$ & 0.26 & 0.65 & 16.90 \\ \hline
LS220* & 715 & 1.00 & $1.20 \times 10^{-7}$ & 0.36 & 0.61 & 47.76 \\
LS220 & 715 & 1.00 & $3.40 \times 10^{-8}$ & 0.26 & 0.61 & 70.38 \\
LS220 & 714 & 1.16 & $1.17 \times 10^{-6}$ & 0.28 & 0.62 & 43.90 \\
LS220* & 714 & 1.16 & $4.33 \times 10^{-6}$ & 0.28 & 0.63 & 36.21 \\
LS220* & 717 & 1.11 & $7.57 \times 10^{-6}$ & 0.35 & 0.66 & 43.25 \\
LS220 & 710 & 1.43 & $1.01 \times 10^{-5}$ & 0.37 & 0.66 & 76.98 \\
LS220 & 707 & 1.66 & $8.39 \times 10^{-5}$ & 0.38 & 0.73 & 52.94 \\ \hline
SFHo* & 413 & 1.00 & $3.97 \times 10^{-5}$ & 0.31 & 0.67 & 52.76 \\
SFHo & 413 & 1.00 & $2.92 \times 10^{-5}$ & 0.30 & 0.66 & 53.13 \\
SFHo* & 412 & 1.13 & $8.40 \times 10^{-5}$ & 0.23 & 0.69 & 25.36 \\
SFHo & 412 & 1.13 & $4.54 \times 10^{-5}$ & 0.29 & 0.67 & 31.69 \\ \hline
SLy4* & 402 & 1.00 & $5.13 \times 10^{-5}$ & 0.31 & 0.67 & 48.85 \\
SLy4 & 402 & 1.00 & $3.21 \times 10^{-5}$ & 0.30 & 0.69 & 44.02 \\
SLy4* & 402 & 1.13 & $1.70 \times 10^{-4}$ & 0.20 & 0.67 & 24.02 \\
\hline\hline
\end{tabular}
\end{center}
\end{table}

When \acp{NS} collide and merger, matter is ejected through a number of 
different physical processes, gaining enough energy to become graviationally 
unbound. 
In particular, the matter ejected within a few dynamical timescales (\ie, ${\sim}10\,$ms) after merger by tidal torques and hydrodynamics shocks driven by core bounces is called dynamical ejecta.
It was found that, within the velocity distribution of the dynamical ejecta, 
some simulations contain also a very fast tail with $\upsilon_{\text{ej}}\geq0.6\,c$ 
\citep{Piran:2012wd,Hotokezaka:2013b,Kyutoku:2012fv,Metzger:2014yda,Ishii:2018yjg,Hotokezaka:2018gmo,Radice:2018pdn}.

The extensive analysis of this tail and its origin in a sample of \ac{NR} 
simulations showed that 
the total mass of this tail dependents on the binary parameters and on the \ac{NS} \ac{EOS}, but it is typically $\sim 10^{-6}-10^{-5} M_{\odot}$. 
This fast tail can be decomposed into two components:
the early fast ejecta that are channeled to high latitudes and that originate at the collisional interface of, predominantly, equal mass models with 
soft \acp{EOS}; and the late fast ejecta, that are largely confined to the plane of the binary, 
and are driven by the shock breakout from the ejecta after the first core bounce \citep{Radice:2018pdn}.

\begin{figure}
    \centering 
    \includegraphics[width=0.49\textwidth]{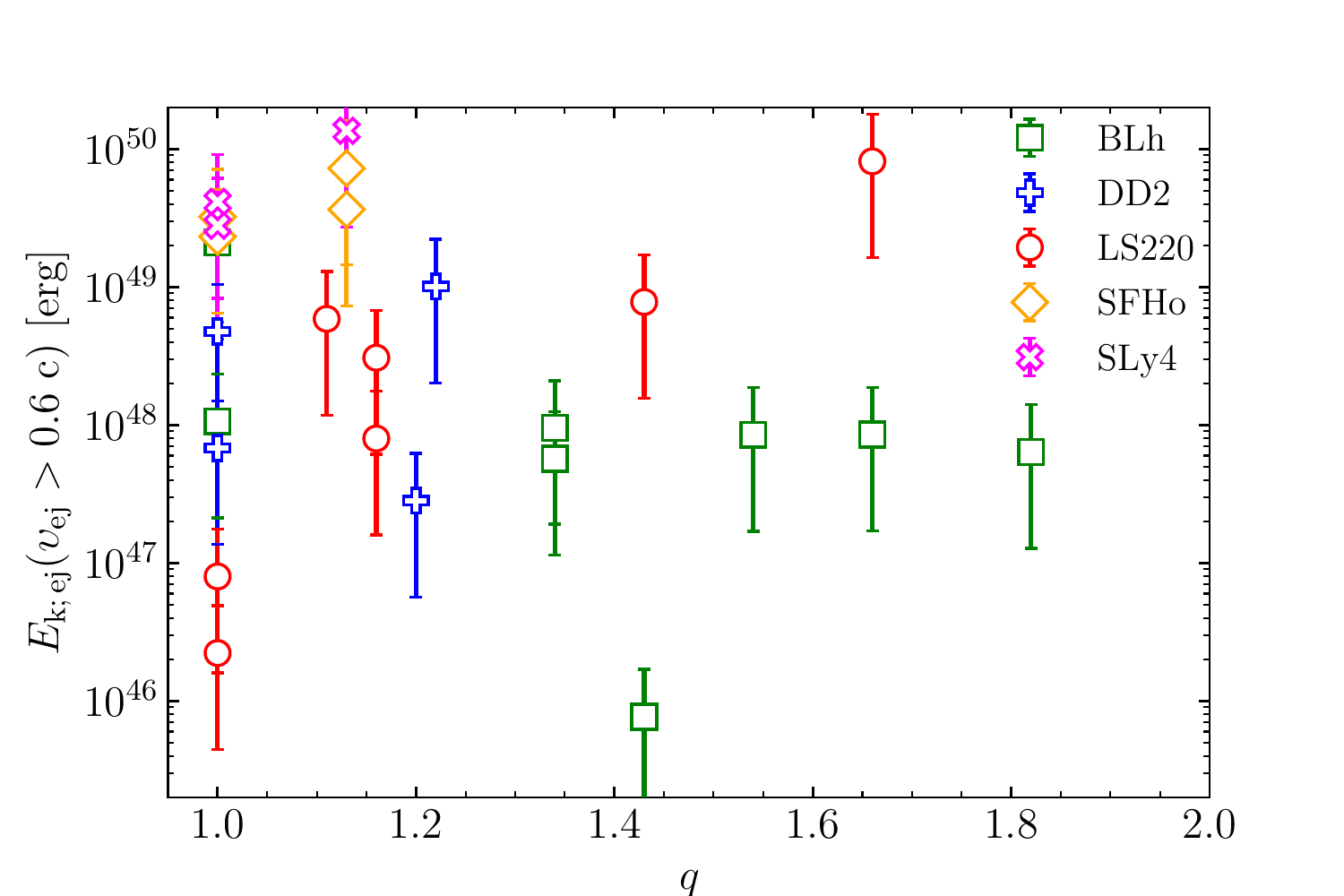}
    \includegraphics[width=0.49\textwidth]{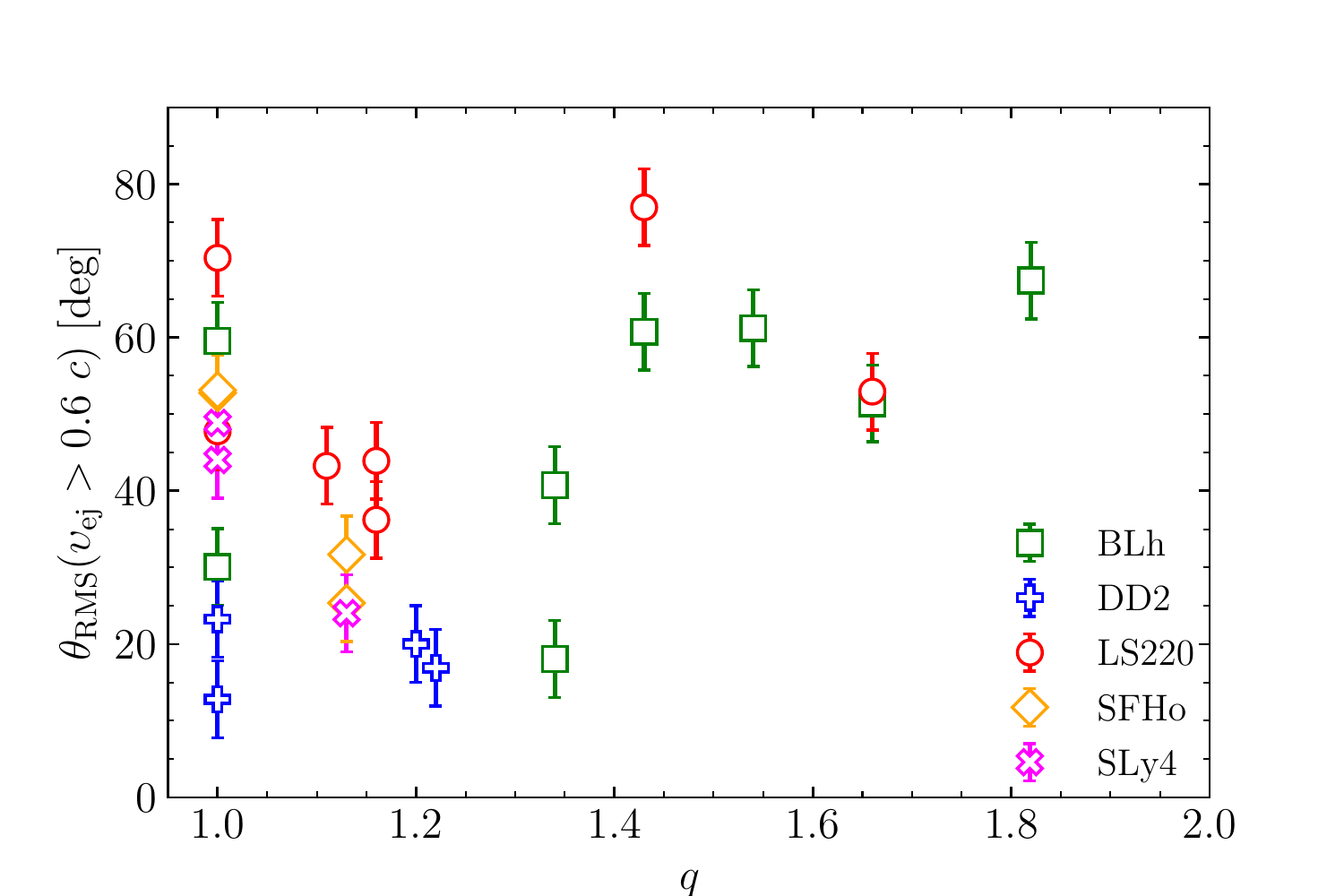}
    \caption{
        Properties of the fast tail of the dynamical ejecta:
        total kinetic energy (\textit{top panel}) and 
        half-\ac{RMS} angle around the binary plane (\textit{bottom panel})
        from a selected set of simulations where this tail is present (see text).
        We assume conservative uncertainties for the angle, $0.5$~deg, 
        and half of the value for the kinetic energy. 
        The top panel shows that only for some \acp{EOS} the 
        total kinetic energy appear to depend on mass ration. Specifically, LS220m SFho and SLy4 \acp{EOS}.
        The half-\ac{RMS} angle appears to depend more on \ac{EOS}, and to be overall larger for high-$q$ models.
    } 
    \label{fig:ejecta_v06}
\end{figure}

\begin{figure}
    \centering 
    \includegraphics[width=0.49\textwidth]{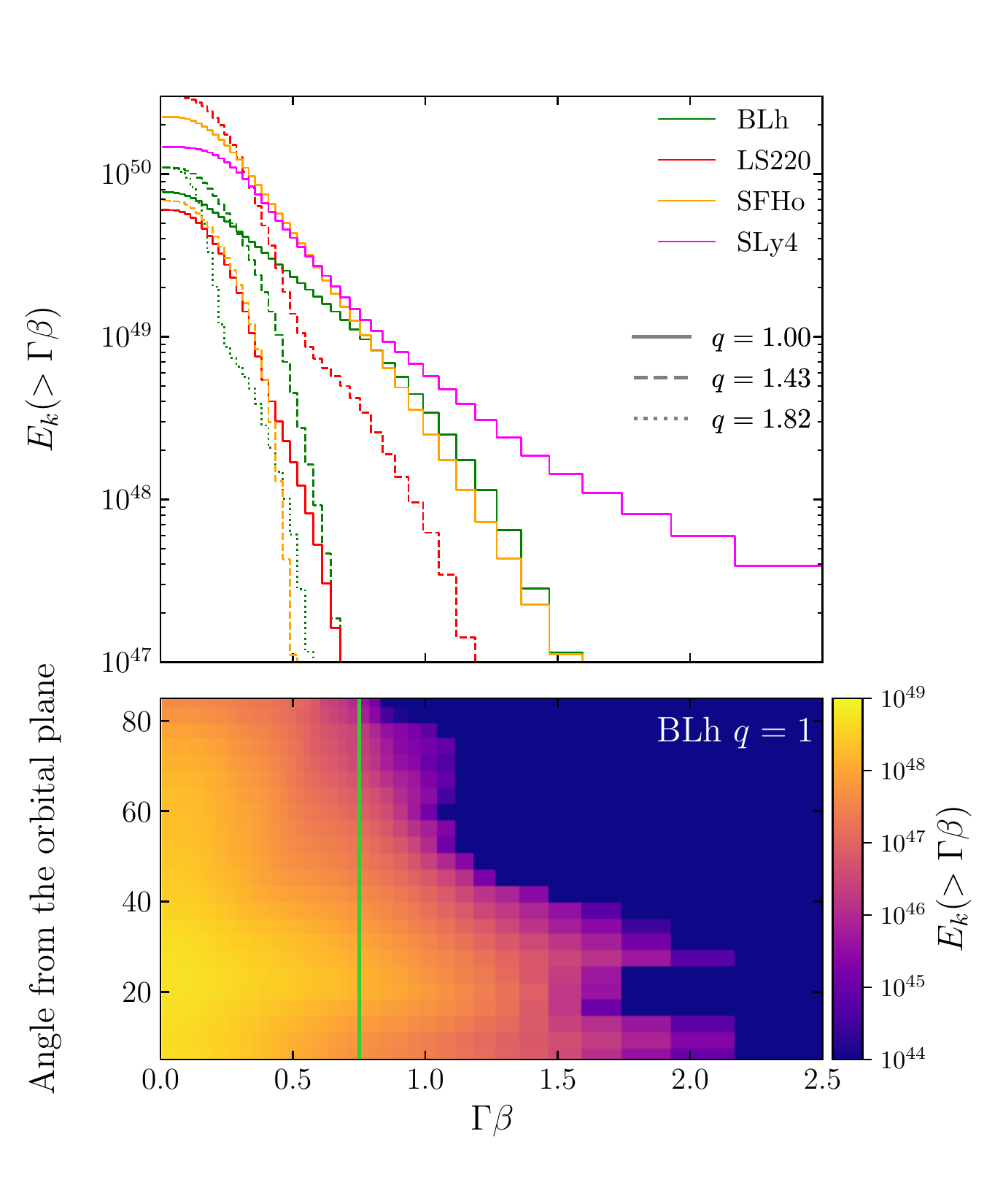}
    \caption{
        Cumulative kinetic energy distribution for a selected set of models (\textit{top panel}) 
        and its angular distribution for a BLh $q=1.00$ model (\textit{bottom panel}).
        The vertical light green line marks the $\upsilon_{\text{ej}}=0.6$.
        The top panel shows that equal mass models have a more extend high energy tail,
        while the bottom panel shows that the angular distribution of the ejecta is not 
        uniform.
    } 
    \label{fig:ejecta_vel_hist}
\end{figure}

In the following we consider the fast ejecta tail in the set of \GW{} targeted simulations (see above). 
The ejecta properties are extracted from the simulations at a coordinate radius of $R=300G/c^2M_{\odot}\approxeq443$~km from the center 
corresponding to the furthest extraction radius available. 
This ensures that the ejecta had the longest possible evolution inside the computational domain. 
This is also consistent with \citet{Radice:2018pdn}.
The simulations were performed at a standard resolution 
with a grid spacing at the most refined grid level $\Delta x \approx 178$~m.
We also performed several simulations with higher resolution,
$\Delta x \approx 123$~m, to asses the resolution effects on ejecta properties.

Notably, not all simulations are found to host a measurable amount of fast ejecta.
Specifically, we find the absence of the fast ejecta component in
simulations with stiff \ac{EOS} and relatively high \mr{}, $q=M_1/M_2\geq1$, where $M_1$ and $M_2$ are the gravitational masses at infinity of the primary and secondary \acp{NS} respectively.
The dynamical ejecta velocity distribution from these models shows 
a sharp cut-off at $\leq0.5c$. 
The absence of fast ejecta can be understood from the fact that 
at large $q$ the ejecta are dominated by the tidal component, whose speed is largely set by the \acp{NS} velocities at 
the last orbit and the system escape velocity.
Additionally, our models with large \mr{} (with fixed chirp mass) experience prompt collapse with no core bounce \citep{Bernuzzi:2020txg}.

The production mechanism of the fast ejecta tail is shown in Fig.~\ref{fig:ejecta_v06_mech}.
Here, we define the fast ejecta tail to consist of material with asymptotic velocity $\upsilon>0.6$~c, following \citet{Radice:2018pdn}. However, we remark that the choice of the velocity threshold $0.6c$ is mostly conventional. We also remark that the synchrotron light curves are computed using the full velocity structure of the ejecta, so this choice does not have any impact on our results.
We find that in our sample of simulations the ejection of mass with 
velocity $\upsilon>0.6$~c coincides with core \bnc{}s, in agreement 
with previous findings by \citet{Radice:2018pdn}.
In models with moderately soft \ac{EOS} or large \mr{}, \eg, the equal mass BLh \ac{EOS} model or the unequal mass models with softer \ac{EOS}, \eg, SFHo \ac{EOS} model, most of the ejecta originate at the first \bnc{}.
However, in equal mass models with very soft \ac{EOS},
\eg, the equal mass SLy4 \ac{EOS} model, we find that additional mass ejection occurs at the second \bnc{}. 
Notably, while the first-\bnc{} component is generally equatorial, the second-\bnc{} component is more polar. This might be attributed to the increased baryon loading of the equatorial region resulting from the slow bulk of dynamical ejecta and with the disc forming matter.

The presence of the fast tail is robust and is not affected by resolution. 
The mass of the fast tail, $\md(\upsilon>0.6\, c)$, however, does have a resolution dependency,
and we find that $\md(\upsilon>0.6\, c)$ changes by a factor of a few
between simulations at standard and high resolutions.
A larger sample of simulations performed at high resolutions is required to asses this uncertainty more quantitatively. 
The mean value of the fast tail mass is
$\overline{\amd}(\upsilon>0.6\, c) = (2.36 \pm 3.89)\times 10^{-5}\,M_{\odot}\ ,$
where we also report the standard deviation.
Other properties of the fast tail, such as velocity, electron fraction and angular distribution, are more robust with respect to resolution, similarly to what is observed for the total dynamical ejecta \citep{Nedora:2020pak}. 

We report the ejecta properties of simulations performed with standard resolution 
Tab.~\ref{tab:sim}.
We find that for most models, the mass averaged velocity of the fast 
tails, $\vd(\upsilon>0.6\,c)$, is close to $0.6c$ with models with softer \acp{EOS} displaying higher velocities. 
The mass-averaged electron, $\yd(\upsilon>0.6\, c)$ is generally above $0.25$, 
indicating that these ejecta were shock-heated and reprocessed by neutrinos. High 
average electron fraction implies that only weak $r$-process nucleosythesis would 
occur, producing elements up to the $2$nd $r$-process peak \citep{Lippuner:2015gwa}.

The total kinetic energy of the fast tail, $E_{\rm k}(\upsilon > 0.6\, c)$,
is shown in top panel of the Fig.~\ref{fig:ejecta_v06}. 
The error bars cover a conservative ${\sim}1$~order of magnitude, that is  
obtained by considering the resolution dependency of the fast ejecta mass and velocity, and by
assuming the same error measures adopted in \citet{Radice:2018pdn}.
The figure shows that the total kinetic energy of the fast tail ranges between ${\sim}10^{46}\,$erg, and ${\geq}10^{50}\,$erg.
Overall, the kinetic energy of the fast tail does not show a strong dependency on the \ac{EOS}, even if very soft \acp{EOS} (like SLy4 and SFHo) tend to have larger energies. 
The dependency on the \mr{} is more prominent, especially for the SLy4, SFHo and LS220 \acp{EOS},
where for the latter, the $E_{\rm k}(\upsilon > 0.6\, c)$ rises by ${\sim3}\,$ orders
of magnitude between $q=1$ and $q=1.7$. 
Notably, for the BLh \ac{EOS} models, the total kinetic energy does not change with the \mr{}.

In the lower panel of the Fig.~\ref{fig:ejecta_v06} we show the \ac{RMS} 
half-opening angle of the fast ejecta around the orbital plane. We assume a conservative error 
of $5$~degrees, motivated by the comparison with higher resolution simulations.
As the angular distribution of fast ejecta depends on the ejection mechanism, the figure
allows to asses which mechanism dominates in each simulation.
The fast ejecta tail is largely confined to the binary plane for the models with stiff
\ac{EOS}, \eg, DD2 \ac{EOS}, where the core \bnc{} ejection mechanism dominate. 
Meanwhile, in simulations with soft \acp{EOS} and high \mr{}s, the fast ejecta has a more 
uniform angular distribution determined by an interplay between
the core dynamics and finite temperature effects driving shocked outflow.

As the mass of the ejecta fast tail shows resolution dependency, so does its total kinetic energy.
For three models for which the fast ejecta were found in both the standard and high resolution simulations, 
we find that $E_{\rm k \,; ej}(\upsilon_{\text{ej}} > 0.6)$ changes
by at least factor of a few. The ejecta \ac{RMS} half-opening angle about the orbital plane is less resolution dependent and its uncertainty is less than ${\sim}50\%$.

Next we consider the distribution of the cumulative kinetic energy of the ejecta, defined as the kinetic energy of the ejecta whose mass is above a certain speed.
We express it as a function of the $\Gamma \beta$ product, where $\beta$ is the ejecta velocity expressed in units of $c$,
and $\Gamma = 1/\sqrt{1-\beta^2}$ is the Lorentz factor.
We show $E_k(>\Gamma\beta)$ for representative set of models in Fig.~\ref{fig:ejecta_vel_hist}. 
The plot displays that for most models the bulk of the 
kinetic energy is allocated to the low velocity matter, 
\ie{} for $\Gamma\beta\leq0.5$. 
Equal mass models show an extended high velocity tail, 
especially the $q=1.00$ model with SLy4 \ac{EOS}.
The bottom panel of the Fig.~\ref{fig:ejecta_vel_hist} shows the 
cumulative kinetic energy distribution in terms of the $\beta \Gamma$ product and angle from the plane of the binary for the $q=1.00$ model with BLh \ac{EOS}.
The distribution is not uniform with respect to the polar angle.
While the high energy tail extends up to the polar angle, the high velocity 
tail is more confined to the orbital plane.
Notably, since the largest part (in mass) of the ejecta is equatorial it eludes the 
interaction with the \ac{GRB} collimated ejecta and expands into an unshocked \ac{ISM}.
The latter can decrease the \ac{ISM} density and delay the peak of the 
synchotron emission \citep{2020MNRAS.495.4981M}.


\begin{figure*}
    \centering 
    \includegraphics[width=0.48\textwidth]{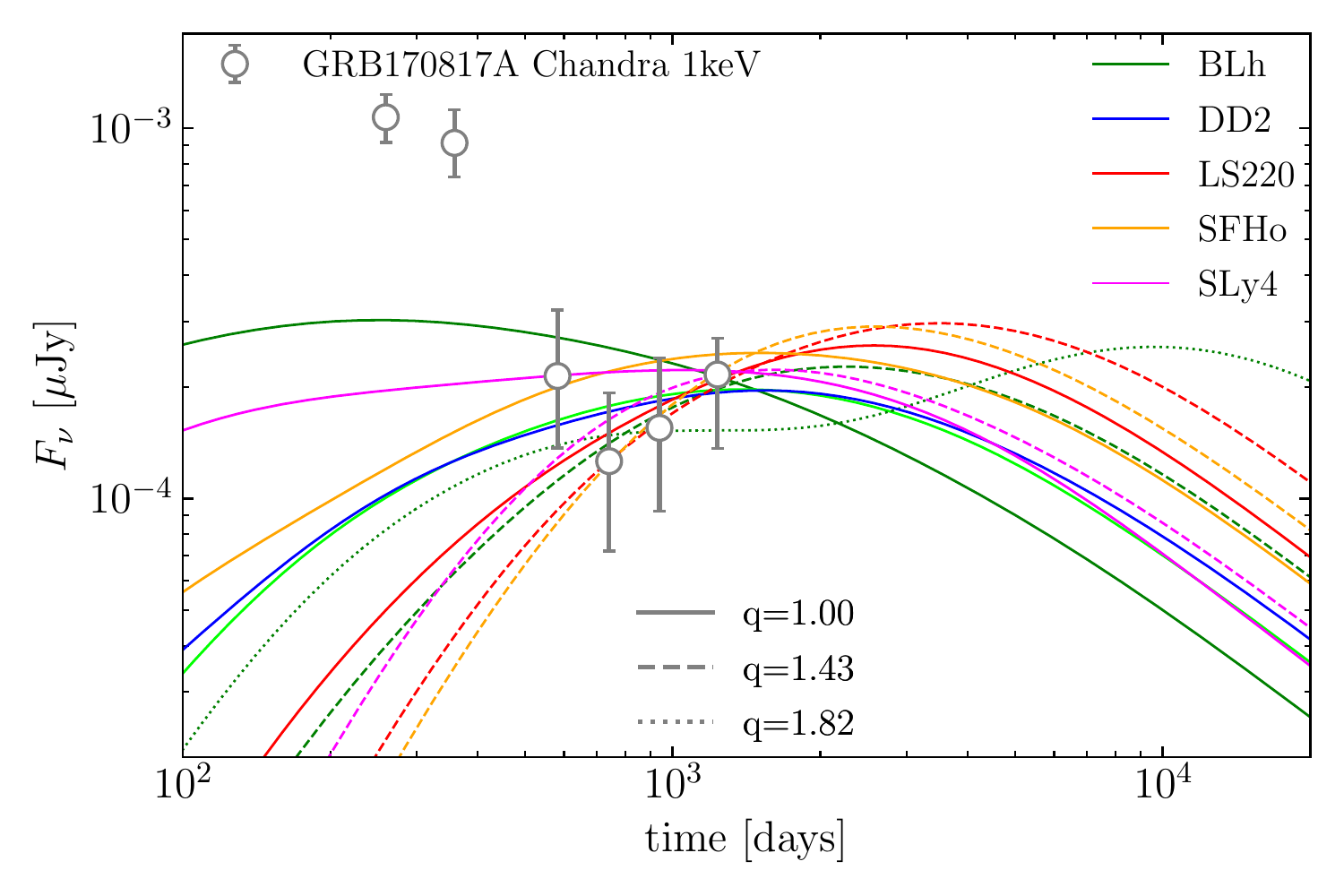}
    \includegraphics[width=0.48\textwidth]{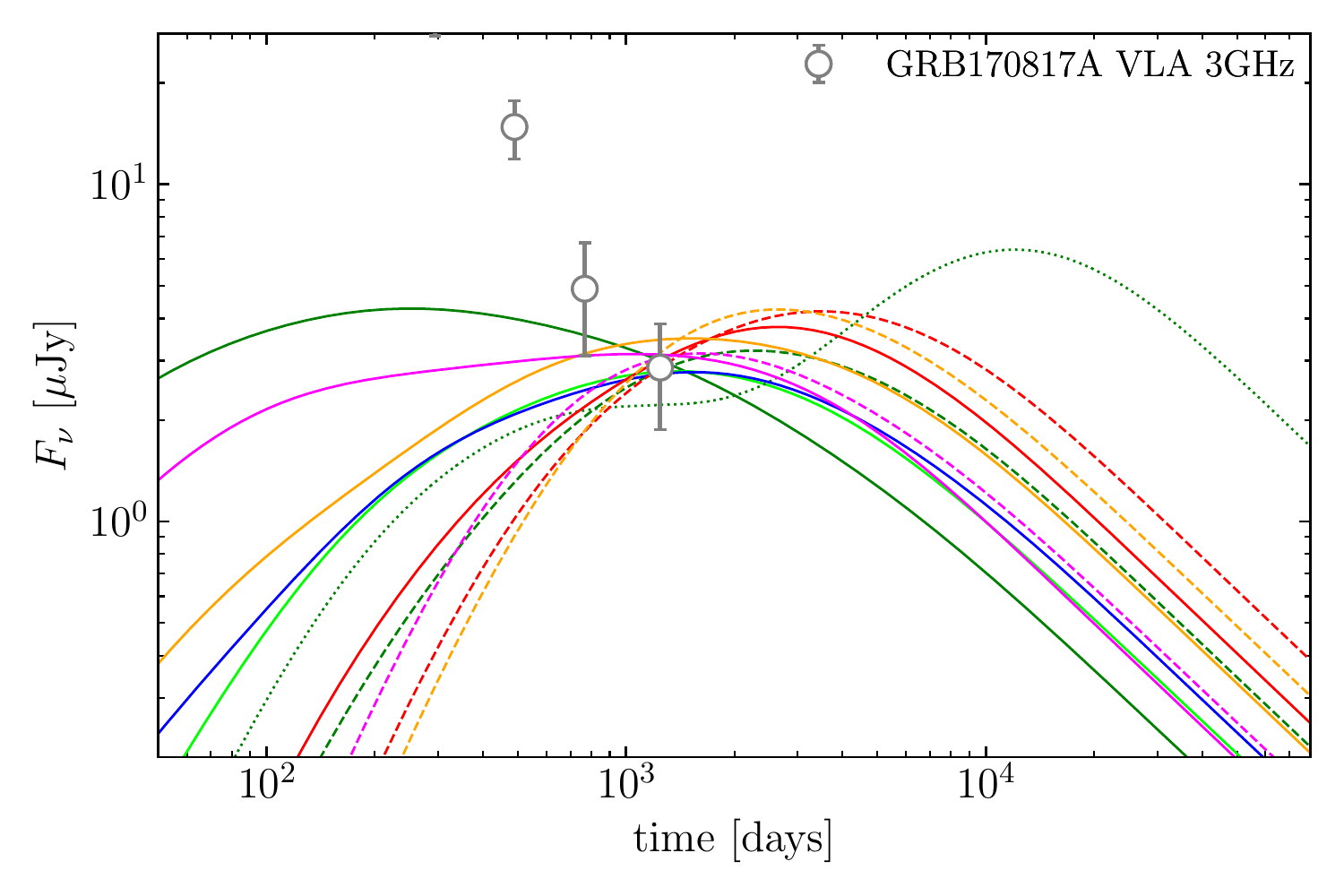}
    \caption{
        Representative kilonova afterglow \acp{LC} for \ac{NR} models, 
        in X-ray (\textit{left panel}) and in radio (\textit{right panel}), where 
        the gray circles are the observational data from \citet{2021arXiv210402070H}
        and \citet{2021arXiv210304821B}.
        The synthetic \acp{LC} are computed with varying 
        micrphysical parameters and \ac{ISM} density within the 
        range of credibility to achieve a better fit to observational data (see Tab.~\ref{tab:pars} for details).
        The plots show that, within allowed parameter ranges, the \acp{LC} 
        from all models are in agreement with observations. 
        Models with moderately stiff \ac{EOS} and $q<1<1.8$ are tentatively preferred,
        as their flux is rising at $t\geq10^3$~days, in agreement with observations.
    } 
    \label{fig:lightcurves}
\end{figure*}

\section{The synchrotron emission from ejecta-\ac{ISM} interaction}

\def\eq{\text{equation}}
\def\eqs{\text{equations}}

Evaluating the synchrotron emission from the merger ejecta requires the calculation of 
the dynamical evolution of the blast wave as it propagates through the \ac{ISM}. 
The dynamical evolution of a decelerating adiabatic blast wave can be described via the 
self-similar solutions. If the blast wave remains always relativistic, the \ac{BM} solution 
\citep{Blandford:1976} applies. If the blast wave remains always subrelativistic, the \ac{ST} 
solution \cite{Sedov:1959} can be used.
Another approach to compute the dynamics of the blast wave is to consider the  
hydrodynamical properties of the fluid behind the shock to be uniform within a given (thin) shell
\citep[\eg][]{Peer:2012,Nava:2013}.
This thin homogeneous shell approximation allows to describe the entire evolution of the shell's 
Lorentz factor from the free coasting phase (where the blast wave 
velocity remains constant) to the subrelativistic phase.
However, there are limitations to this approach. Specifically, it was shown 
to differ from \ac{BM} self-similar solution in the ultrarelativistic regime by a numerical factor \citep{Panaitescu:2000bk},
and a self-similar solution of the non-relativistic deceleration \citep{Huang:1999di}.
In application to the mildly relativist ejecta with velocity structure, 
the deviation was shown to be of order unity 
\citep{Piran:2012wd,Hotokezaka:2015eja}.

We calculate the non-thermal 
radiation arising from the dynamical ejecta propagating into the cold \ac{ISM} 
with the semi-analytic code \texttt{PyBlastAfterglow}.
The method can be summarized as following. 
For a given distribution of energies as a function of velocity, 
we divide the ejecta into velocity shells and solve the adiabatic radial expansion of
the ejecta in the thin shell approximation at each polar angle using the kinetic energy 
distributions discussed in Sec.~\ref{sec:ejecta}.
See also \eg, \citet{Piran:2012wd} and \citet{Hotokezaka:2015eja} for similar treatments.

For the adiabatic evolution we adopt the blast wave dynamics formalism developed by 
\citet{Nava:2013} where the evolution of the blast wave Lorentz factor is given by their \eqs{}~3-7,
which we solve numerically via a 4th order adaptive step \ac{RK} method.
We neglect the effects of radiation losses and lateral spreading of the blast wave and 
focus on its evolution prior to and shortly after the onset of the deceleration.
The \ac{EOS} assumed is that of the ideal transrealtivitisc fluid, where the adiabatic 
index is given as a function of the normalized temperature \citep[\eq~11 in][]{Peer:2012} which is
computed adopting the polynomial fit \citep[\eq~5 in][]{Service:1986}

Next, we consider the forward shock propagating into the upstream medium as the blast wave expands.
The bulk of the energy is being deposited into the non-thermal protons. Part of this 
energy is transferred to relativistic electrons via complex shock interactions. 
It is however possible to consider a simplified prescription for the transfer of energy from 
protons to electrons \citep[\eg][]{Dermer:1997pv}. 
A fraction $\varepsilon_e$ and $\varepsilon_B$ of shock internal energy is assumed to be deposited 
into the relativistic electrons and magnetic field respectively.
The injected electrons are assumed to have a power-law distribution 
$dN/d\gamma_e\propto\gamma^{-p}$, where $\gamma_e$ is the electron Lorentz factor,
$p$ is the spectral index, a free parameter.
The critical Lorentz factors of the spectrum are the minimum one, $\gamma_{\rm min}$, and the critical one, $\gamma_c$, computed via standard expressions 
\citep[\eqs~A3 and A4 in][respectively]{Johannesson:2006zs}.
Depending on the ordering of the $\gamma_{\rm min}$ and $\gamma_c$, two regimes are considered,
namely the \textit{fast cooling} regime if $\gamma_{\rm min}>\gamma_c$, and \textit{slow cooling}
regime otherwise \citep{Sari:1997qe}.

\begin{table}
    \begin{center}
    \caption{
        List of parameters for synthetic \acp{LC} shown in the Fig~\ref{fig:lightcurves} 
        and Fig.~\ref{fig:lightcurve_peaks}.
        For the former the microphysical and \ac{ISM} density are adjusted model-wise 
        to achieved the good agreement with observations. For the latter,
        (the last row of the table) the parameters are the same for all models shown.
        Other parameters, such as observational angle, are the same everywhere (see text).
    }
    \begin{tabular}{l | l l l l}
    Fig~\ref{fig:lightcurves} & $p$ & $\epsilon_e$ & $\epsilon_b$ & $n_{\text{ISM}}$ \\ \hline 
    BLh q=1.00    & 2.05 & 0.1          & 0.002        & 0.005            \\
    BLh q=1.43    & 2.05 & 0.1          & 0.003        & 0.005            \\
    BLh q=1.82    & 2.05 & 0.1          & 0.01         & 0.01             \\
    DD2 q=1.00    & 2.05 & 0.1          & 0.005        & 0.005            \\
    LS220 q=1.00  & 2.05 & 0.1          & 0.01         & 0.005            \\
    LS220 q=1.43  & 2.05 & 0.1          & 0.001        & 0.005            \\
    SFHo q=1.00   & 2.05 & 0.1          & 0.001        & 0.004            \\
    SFHo q=1.43   & 2.05 & 0.1          & 0.01         & 0.005            \\
    SLy4 q=1.00   & 2.05 & 0.1          & 0.001        & 0.004            \\
    SLy4 q=1.43   & 2.05 & 0.1          & 0.004        & 0.005            \\ \hline
    Fig.~\ref{fig:lightcurve_peaks}  & 2.15 & 0.2         & 0.005        & 0.005           
    \end{tabular}
    \end{center}
    \label{tab:pars}
\end{table}

The comoving synchrotron \ac{SED} is approximated with a smooth broken power 
law according to \citet{Johannesson:2006zs}, and computed with their \eqs~A1 and A7 for the slow cooling regime and their \eqs~A2 and A6 for the fast cooling regime. 
The characteristic frequencies are obtained from the characteristic Lorentz factors 
$\gamma_{\rm min}$ and $\gamma_c$ via their \eq~A5.

The synchrotron self-absorption is included via flux attenuation \citep[\eg][]{Dermer:2009}. %
However, for the applications discussed in this paper, the self-absorption is not 
relevant as the ejecta remains optically thin for the emission ${\geq3}\,$GHz 
\citep[\eg][]{Piran:2012wd}.

We compute the observed flux, integrating over the \ac{EATS}, following 
\citet{Lamb:2018ohw}.
For each segment of the blast wave, the time for the observer is evaluated via their 
\eq~3, and then the observed, Doppler-shifted flux is obtained via their 
\eq~2.
See \citet{Salmonson:2003} for the detailed discussion of the method 
and \citet{2021arXiv210402070H} for a similar implementation.

In the ultrarelativistic regime, evolving a single velocity shell,
the code was found to be consistent with \texttt{afterglowpy} \citep{Ryan:2019fhz}, 
while in the subrelativistic regime, modeling the kilonova
afterglow, the code produces \acp{LC} consistent with 
the model of \citet{Hotokezaka:2015eja},
which was applied to the \ac{BNS} ejecta in \citet{Radice:2018pdn}.

It is however important to note that the methods discussed above for both the blast wave dynamics and the synchrotron emission become increasingly inaccurate as the 
blast wave decelerates and spreads, and as most of the electrons become subrelativistic.
So we do not discuss the late-time emission after the \ac{LC} peak.
We discuss different physics implemented in \texttt{PyBlastAfterglow} with application 
to both structured \acp{GRB} and analytic ejecta profiles elsewhere.

The free parameters of the model are chosen as follows.
We assume the \ac{ISM} density to be uniform within the range 
$\nism\in(10^{-3}, 10^{-2})\,\ccm$ \citep{Hajela:2019mjy}. 
The observational angle, defined as the angle between the line of sight and the polar axis of the \ac{BNS} system, is $\theta_{\text{obs}}=30$~deg \citep{TheLIGOScientific:2017qsa}.
For the luminosity distance of NGC 4993, the host galaxy of \GW{},
we adopt $41.3\times10^{6}$~pc with the redshift $z=0.0099$ \citep{Hjorth:2017yza}.

Recent Chandra observations showed that the emerging component
in \GRB{} afterglow is accompanied by the onset of the spectral evolution
\citep{2021arXiv210402070H}. 
The observation data analysis suggests that a lower value for the electron power law distribution slope is more favorable, $p=2.05$, but at low significance level.
Due to this uncertainties, 
we consider the following parameter ranges
$\varepsilon_e\in(0.1, 0.2)$,
$\varepsilon_B\in(10^{-3}, 10^{-2})$, 
$p\in[2.05,2.15]$.
Additionally, the effect of a lower $p=2.05$ is discussed in \citet{2021arXiv210402070H}.

\section{Results}

We show X-ray and radio \acp{LC} from several representative models 
in Fig.~\ref{fig:lightcurves}, alongside the latest \GRB{} observational data.

The \ac{LC} shape is determined by the ejecta velocity and angular distribution.
For instance, models with broad velocity distribution, such as equal mass model 
with SLy4 \ac{EOS} (see Fig.~\ref{fig:ejecta_vel_hist}), have a wide \ac{LC}. 
This \ac{LC} start to rise very early (in comparison with the high \mr{} model) 
as the fast velocity shells decelerate 
and peak on a shorter timescales.
However, models with with narrower velocity distribution, 
such as the model with LS220 \ac{EOS} and $q=1.43$, have a narrower \ac{LC} that rises later (${\sim}10^2$~days after merger).

Generally, within the uncertain microphsycis and \ac{ISM} density, the kilonova afterglow from most models is in a good agreement with the 
new observations by Chandra.
In particular, models with $1.00<q<1.82$ and moderately stiff \acp{EOS} show rise in flux  $\geq10^3$~days postmerger, in agreement with observations

Fixing the microphysical parameters and \ac{ISM} density to 
$\nism=5\times10^{-3}~\gccm{}$, $\epsilon_e=0.1$ and $\epsilon_b=5\times10^{-3}$, 
we find that the flux at the \ac{LC} peak, $F_{\nu;p}$, 
is the largest for soft \acp{EOS} such as SFHo and SLy4.
Overall, however, the peak flux seems largely independent of \ac{EOS}.
The $F_{\nu;p}$ dependency on the \mr{} for soft \acp{EOS} is that, 
the higher the \mr{}, the lower the peak flux. 
This can be understood from the following considerations.
While the ejecta total kinetic energy budget of these models increases 
with the \mr{}, the mass-averaged velocity decrease 
(see Fig.~$5$ in \citet{Nedora:2020pak}).
And slower, more massive ejecta have lower peak flux.
However, for stiffer \acp{EOS}, such as DD2 and LS220, the $q$ dependency is less clear.

The \ac{LC} shape and the peak time depend weakly on the uncertain 
microphysics and $n_{\text{ISM}}$. Specifically, within $n_{\text{ISM}}\in(10^{-3},10^{-2})~\ccm$, $t_p$ varies by a factor of a few.
An additional source of uncertainties is the ejecta properties dependency on numerical
resolution. However we estimate its effect on the \acp{LC} to be smaller than
that of the unconstrained shock microphsyics and $n_{\text{ISM}}$.
Specifically, the $t_p$ changes by a factor of ${\leq2}$, 
and $F_{\nu;p}$ chages withing a factor of ${\leq4}$.

\section{Discussion}

\begin{figure}
    \centering 
    \includegraphics[width=0.49\textwidth]{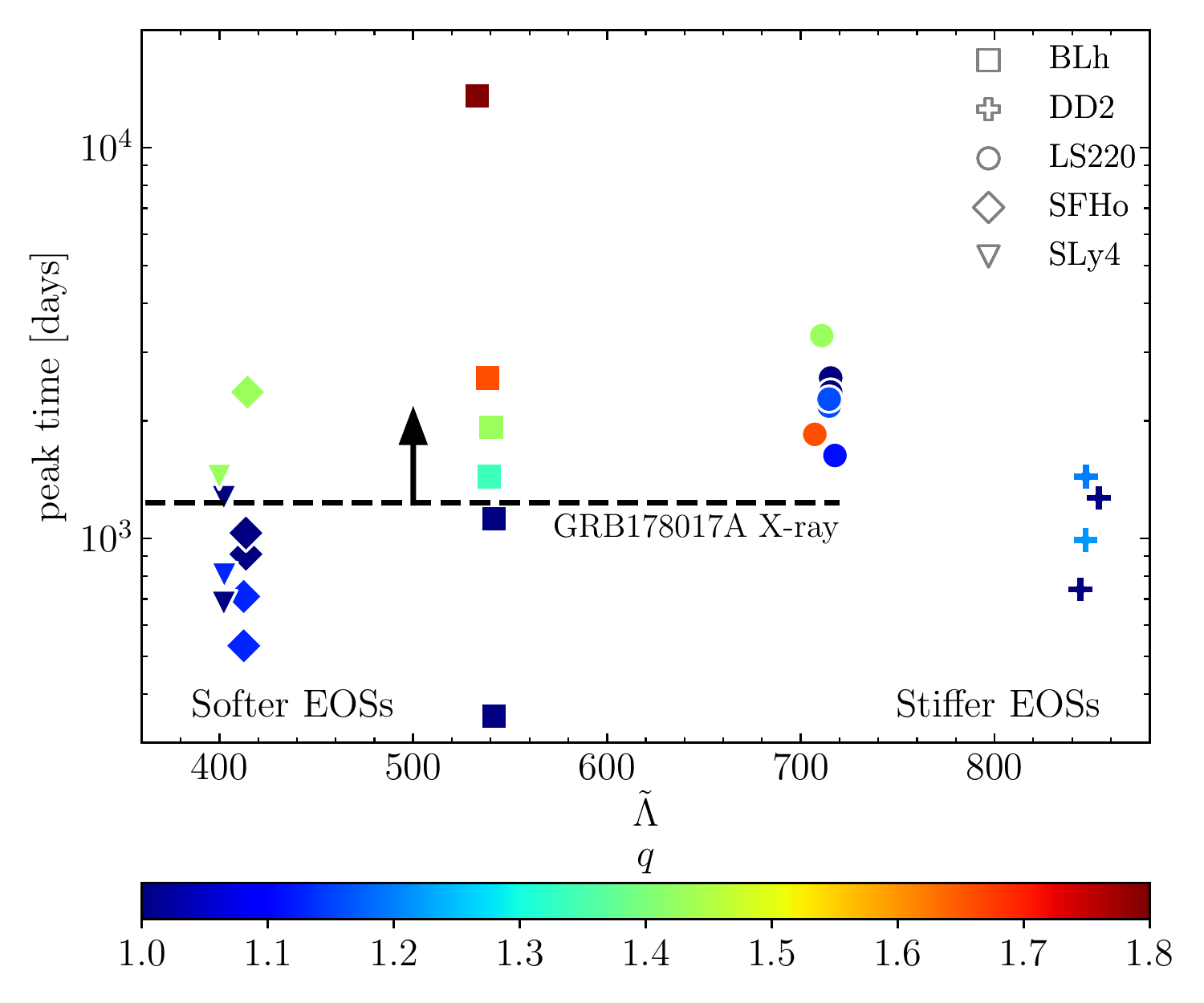}
    \caption{
        Peak time, $t_p$, for \ac{LC} for all considered \ac{NR} simulations. 
        Dashed black line corresponds to the last observation of \GRB{} afterglow,
        where the rising flux implies that it is a lower limit on the kilonova 
        afterglow.
        The microphysical parameters and \ac{ISM} density for all models are fixed and 
        given in the Tab.~\ref{tab:pars}.
        The plot shows that in general the $t_p$ increases with mass ration and with 
        softness of the \ac{EOS}, except for the softest, DD2 \ac{EOS}. 
    } 
    \label{fig:lightcurve_peaks}
\end{figure}

In this work we analyzed a large set of \ac{NR} \ac{BNS} simulations performed with 
state-of-the-art \ac{NR} code \texttt{WhiskyTHC} and targeted to \GW{}.
Simulations included the effects of neutrino emission and reabsorption, 
effective viscosity via subgrid turbulence, and microphysical finite-temperature 
\acp{EOS}.

We found that most simulations' ejecta contains material with velocities $\gtrsim 0.6$~c, 
whose properties are in agreement with previous studies \citep{Radice:2018pdn}.
However, in binaries with large \mr{} and/or prompt BH formation 
(that experience weaker/no core bounce at merger), the fast ejecta could be absent.

The latest \GRB{} observations by Chandra at $10^3$~days showed a changing afterglow behaviour in X-ray band \hajela{}. 
We suggest that this change can be attributed to the emergence of the kilonova afterglow.
In particular, we evolved the ejecta from \ac{NR} simulations with a semi-analytic 
code and computed its synchrotron emission. 
We found that the synthetic \acp{LC} are in agreement with the emerging new component 
in the \GRB{} afterglow within the range of credibility of the microphisycal parameters 
and of the \ac{ISM} density, $n_{\text{ISM}}$. 
Additionally, the change in \GRB{} afterglow has the following implications: 
the kilonova afterglow peak should be (i) later and (ii) brighter than 
the latest observations.
The condition (ii) is not particularly strong, as the \ac{LC} peak flux 
depends sensibly on the uncertain microphysical parameters.
The (i) condition is more robust from that point of view and allows to gauge
which \ac{NR} models from our sample have afterglow predictions more supported by observations.

In Fig.~\ref{fig:lightcurve_peaks} we show the time of the \ac{LC} peak 
for all models (for fixed microphysics and $n_{\text{ISM}}$, 
see Tab.~\ref{tab:pars}), including those with absent fast tail, as 
these models still have sufficient amount of mildly relativistic material to produce a bright afterglow.
The time of the synchrotron \ac{LC} peak, $t_p$, of all the models 
is distributed around ${\sim}10^3$~days postmerger (except tidal disruption cases, 
as the model with $q=1.82$ and BLh \ac{EOS}). 
Models with small \mr{} tend to have $t_{p}<10^3$~days, while 
more asymmetric models point towards $t_p>10^3$~days.
This trend is more apparent for models with soft \acp{EOS}. 
The reason for this lies in the mechanism responsible for the the fast ejecta tail
(see Sec.~\ref{sec:ejecta}).
As the \mr{} increases, the amount of the shocked ejecta component decreases,
and so does the kinetic energy of the ejecta fast tail. 
The afterglow of the slower, more massive ejecta peaks later \citep[\eg][]{Hotokezaka:2015eja}.
Indeed, the time of the \ac{LC} peak depends primarily on the ejecta 
dynamics, the so-called deceleration time \citep[\eg][]{Piran:2012wd}.
Additionally, in Fig.~\ref{fig:lightcurve_peaks} we show the lower limit on $t_p$, 
(the time of the latest observation).
We observe that \acp{LC} of models with moderate amount of fast ejecta, 
\eg, asymmetric models with \acp{EOS} of mild stiffness, lie above the limit, 
while models with highly energetic fast tails, such as equal mass models with very stiff 
\acp{EOS}, peak too early. 
It would be interesting to combine these constraints with those obtained from the modeling of the thermal component of the kilonova. However, this is not presently possible because the thermal emission is expected to be dominated by slowly expanding secular winds from the merger remnant, which cannot be presently simulated in full-NR \citep[see \eg][]{Nedora:2019jhl}

The main conclusion of our work is that the observed \GRB{} changing behaviour at $10^3$~days after 
merger can be explained by the kilonova afterglow produced by ejecta in ab-initio \ac{NR} \ac{BNS} 
simulations targeted to \GW{}. 
Specifically, models that produce a mild amount of fast ejecta, those with 
moderately to large \mr{} and moderately stiff \acp{EOS} are favoured.
The dominant uncertainties in our analysis are the ill-constrained microphysical parameters of 
the shock. Additionally, the systematic effects due to the finite resolution, neutrino treatment 
and \acp{EOS} might be important. 
A larger set of observations, that allows for a better assessment of shock microphysics, and a
larger sample of high resolution \ac{NR} simulations are required to investigate these uncertainties further. 
We leave this to future works.

\section*{Acknowledgements}
DR acknowledges support from the U.S. Department of Energy, Office of Science, Division of Nuclear Physics under Award Number(s) DE-SC0021177 and from the National Science Foundation under Grant No. PHY-2011725.
S.B. and B.D. acknowledge support by the EU H2020 under ERC Starting Grant, no.~BinGraSp-714626.
Numerical relativity simulations were performed on the supercomputer
SuperMUC at the LRZ Munich (Gauss project pn56zo), on supercomputer Marconi at CINECA (ISCRA-B project numberHP10BMHFQQ); on the supercomputers Bridges, Comet, and Stampede (NSF XSEDE allocation TG-PHY160025); on NSF/NCSA Blue Waters (NSF AWD-1811236); on ARA cluster at Jena FSU.
AP acknowledges PRACE for awarding access to Joliot-Curie at GENCI@CEA, France (project ra5202).
This research used resources of the National Energy Research Scientific Computing Center, a DOE Office of Science User Facility supported by the Office of Science of the U.S.~Department of Energy under Contract No.~DE-AC02-05CH11231.
Data postprocessing was performed on the Virgo ``Tullio'' server at Torino supported by INFN.
The authors gratefully acknowledge the Gauss Centre for Supercomputing e.V. (\url{www.gauss-centre.eu}) for funding this project by providing computing time on the GCS Supercomputer SuperMUC at Leibniz Supercomputing Centre (\url{www.lrz.de}). 

\textit{Software:} We are grateful to the countless developers contributing to open source projects on which
this work relied including \texttt{NumPy} \citep{numpy}, \texttt{Matplotlib}, \cite{matplotlib} and \texttt{SciPy} \cite{scipy}.

\textit{Data Availability:} the lightcurves and table \ref{tab:sim}
are available at \cite{nedora_vsevolod_2021_4700060}.

\bibliography{refs20210628}

\end{document}